\documentclass{article}

\usepackage{PRIMEarxiv}

\usepackage[utf8]{inputenc} % allow utf-8 input
\usepackage{CJKutf8}        % Chinese probe text under pdfLaTeX
\usepackage[T1]{fontenc}    % use 8-bit T1 fonts
\usepackage{hyperref}       % hyperlinks
\usepackage{url}            % simple URL typesetting
\usepackage{booktabs}       % professional-quality tables
\usepackage{amsfonts}       % blackboard math symbols
\usepackage{amssymb}        % additional symbols, including \varnothing
\usepackage{amsmath}
\usepackage[numbers]{natbib} % numeric citations compatible with unsrt
\usepackage{nicefrac}       % compact symbols for 1/2, etc.
\usepackage{microtype}      % microtypography
\usepackage{lipsum}
\usepackage{fancyhdr}       % header
\usepackage{graphicx}       % graphics
\usepackage{subcaption}     % subfigures
\usepackage{wrapfig}        % wrapped figures
\usepackage{enumitem}       % customized research-question lists
\usepackage{tabularray}     % fixed-width appendix tables
\UseTblrLibrary{booktabs}
\usepackage[ruled,vlined,linesnumbered]{algorithm2e}
\SetKwInput{KwOracle}{Oracles}
\SetKwInput{KwIn}{Inputs}
\usepackage{cleveref}
\usepackage{color}
\graphicspath{{media/}}     % organize your images and other figures under media/ folder

\newif\ifdraft
\drafttrue
\ifdraft
  \newcommand{\todocolor}[2]{{\color{#1}#2}}
\else
  \newcommand{\todocolor}[2]{}
\fi

\newcommand{\figref}[1]{Figure~\ref{#1}}

\newcommand{\tabref}[1]{Table~\ref{#1}}

\newcommand{\secref}[1]{Section~\ref{#1}}

\definecolor{algxrefgray}{gray}{0.55}

\newcommand{\eqnref}[1]{Equation~\ref{#1}}
\renewcommand{\eqref}[1]{\eqnref{#1}}

\newcommand{\method}[0]{DFA\xspace}

\title{Is Your LLM Relay Serving What It Claims?\\ A low-cost and high-power test for LLM Relay Auditing
}

\author{
  Fangran Miao\thanks{Co-first authors} \\
  Department of Data Science and Artificial Intelligence \\
  The Hong Kong Polytechnic University \\
  Hong Kong SAR\\
  \texttt{fangran.miao@connect.polyu.hk} \\
    \And
  Shengjie Niu$^*$ \\
  Department of Data Science and Artificial Intelligence \\
  The Hong Kong Polytechnic University \\
  Hong Kong SAR\\
  \texttt{shengjie.niu@connect.polyu.hk} \\
    \AND
  Jian Huang\thanks{Corresponding authors} \\
  Department of Data Science and Artificial Intelligence \\
  The Hong Kong Polytechnic University \\
  Hong Kong SAR\\
  \texttt{j.huang@connect.polyu.hk} \\
    \And
  Ting Li\footnotemark[2] \\
  Department of Statistics and Data Science \\
  Southern University of Science and Technology \\
  Shenzhen\\
  \texttt{liting@sustech.edu.cn} \\
}

\begin{document}
\maketitle

\begin{abstract}
Relay and reseller APIs have become a common way to access large language models (LLMs), yet users cannot directly verify whether they receive the models they request and pay for. 
This opacity motivates LLM relay auditing, but existing methods struggle to combine low cost, robustness to benign deployment changes, and statistically interpretable decisions under black-box access. 
To address these limitations, we propose distributional fingerprint auditing (DFA), a statistical method that requires only textual responses from suspect and trusted reference endpoints. DFA repeatedly queries semantically simple finite-choice probes and uses the resulting categorical response distributions to construct endpoint fingerprints. It compares the suspect and reference fingerprints probe by probe and aggregates the resulting evidence using the Cauchy combination test, yielding an approximate endpoint-level $p$-value for audit decisions at a prespecified nominal significance level.
Using registered fingerprints of candidate models, a likelihood-based formulation further supports model attribution under full substitution and joint estimation of substitute identity and substitution ratio under constant-ratio mixed routing. Our benchmark includes 16 proprietary and open-weight LLMs from 10 model families. Across 240 ordered full-substitution pairs, DFA achieves high detection rates at a nominal significance level of $\alpha=0.05$, using five probes and 250 relay requests per audit with stored reference fingerprints. Configuration overrides can lead to rejection even when the served model remains unchanged, with effects varying across models and configurations. Simulations based on stored empirical fingerprints demonstrate accurate model attribution and substitution-ratio estimation, with estimation error decreasing as the audit budget increases.
\end{abstract}

\section{Introduction}\label{sec:intro}

Relay and reseller APIs have become a canonical way to access large language models (LLMs)~\citep{vercel2026gateway,openrouter2026quickstart,litellm2026routing}.
These third-party services expose a unified API that aggregates access to multiple underlying models and inference providers.
Users can specify a model identifier in each API request; the relay handles backend routing and returns the generated response.
This convenience, however, comes at the cost of backend opacity: users cannot directly verify whether the returned responses were actually generated by the requested model.
A dishonest relay may silently substitute the advertised model with a cheaper alternative, such as a smaller model or a quantized variant, while continuing to charge users at the advertised model's rate.
Differences in inference costs create a direct incentive for relays to serve cheaper substitutes, while the risk of detection has historically been low~\citep{sun2025invisible,velasco2025your}.
This concern is not merely hypothetical: recent measurement work finds deceptive model claims among shadow APIs at a meaningful scale, with consequences for research reproducibility and service reliability~\citep{zhang2026real}.
More seriously, undisclosed malicious fine-tuning can introduce harmful or backdoored behavior, while hidden jailbreak-enabling system prompts can bypass safety safeguards and induce the model to comply with harmful requests~\citep{egashira2024exploiting,shen2024anything,zhu2026rut}.
These risks motivate \emph{LLM relay auditing}, which first tests whether a relay's responses are consistent with those of its advertised model. Upon detecting an inconsistency, the auditor further seeks to identify the substitute model and estimate the substitution ratio.

Despite its practical importance, sound relay auditing imposes several requirements that existing methods struggle to satisfy simultaneously.
First, auditors are often third parties with only \emph{black-box} query access: they can submit requests and observe textual responses but cannot inspect model internals or access backend execution details. This excludes methods requiring model-owner cooperation or access to reference-model token probabilities~\citep{xu2024instructional,kirchenbauer2023watermark,zhu2026rut}. 
Second, practical audits should be cost-efficient and robust to benign deployment changes. Methods based on full completions or semantic perturbations require longer generations or additional queries, increasing audit costs~\citep{met,shao2025zeroprint}. Moreover, changes in system prompts or decoding settings can substantially shift response distributions even when the served model remains unchanged~\citep{kbf,tee}, potentially causing these methods to falsely flag model substitution.
Third, auditing requires statistically interpretable decisions. Distance-based verification methods compare fingerprint distances against a decision threshold but do not provide a null-hypothesis $p$-value for each audit~\citep{bruckner2026onetoken}. 
These considerations motivate a low-cost, black-box audit that provides statistically interpretable endpoint-level decisions.

In this work, we propose distributional fingerprint auditing (DFA), a low-cost and high-power statistical test for LLM relay auditing.
Specifically, we construct a set of finite-choice probes and repeatedly query each endpoint to estimate its response distribution for each probe.
These distributions capture model-specific behavioral patterns that provide evidence about the underlying model's identity~\citep{pasquini2025llmmap,sun2025idiosyncrasies}.
Together, these distributions form an empirical \emph{distributional fingerprint} that characterizes the endpoint's responses across the probe suite.
To audit a suspect relay, we compare its fingerprint with that of a trusted reference endpoint serving the advertised model.
We test for distributional differences on each probe and combine the resulting probe-specific $p$-values into an endpoint-level statistic using the Cauchy combination test~\citep{cct}.
This aggregation emphasizes strong evidence from small $p$-values, helping preserve evidence of inconsistency when only a subset of probes distinguishes the endpoints.
A Cauchy approximation to the upper tail of the combined statistic's null distribution yields an endpoint-level $p$-value, supporting a statistically interpretable decision on whether to reject distributional consistency between the suspect relay and the trusted reference.

To further characterize a detected inconsistency, we examine two forms of model substitution: full substitution and mixed routing~\citep{kbf}.
Under full substitution, the relay serves a single substitute model for all audited requests.
Under mixed routing, the relay serves the advertised model for some requests and a substitute model for others.
For further inference, the auditor considers a set of candidate substitute models that the relay may serve in place of the advertised model. Given additional black-box access to trusted endpoints serving these candidates, we construct and register their distributional fingerprints using the same probe suite.
Using these fingerprints together with that of the advertised model, we introduce a likelihood-based formulation that evaluates how well a candidate model, or its mixture with the advertised model, explains the suspect relay's observed responses.
Under full substitution, maximizing this likelihood identifies the candidate whose fingerprint best explains the observed responses, providing model attribution.
Under constant-ratio mixed routing, maximizing the likelihood jointly estimates the substitute model and the proportion of audited requests routed to it.

We evaluate DFA through five research questions covering substitution detection, audit cost, robustness to configuration overrides, model attribution, and substitution-ratio estimation. 
Our experiments cover 16 LLMs from 10 model families, including proprietary models such as \texttt{gpt-6-luna} and \texttt{claude-sonnet-5.5}, and open-weight models such as \texttt{gpt-oss-120b} and \texttt{nemotron-3-ultra}.
Across 240 ordered full-substitution pairs, including within-family substitutions, DFA achieves high detection rates at a nominal significance level of $\alpha=0.05$, using five probes and 250 relay requests per audit with stored reference fingerprints.
Across the four models evaluated in the cost analysis, a 250-request audit costs an estimated USD~0.003--0.048, with the stored reference fingerprint reused across audits.
Configuration-override experiments show that even minor changes can lead to rejection for some models, while non-rejection rates vary substantially across models and configurations.
Simulations based on stored empirical fingerprints show high attribution accuracy under full substitution and increasingly precise substitution-ratio estimates under constant-ratio mixed routing as the audit budget increases.

Our contributions are threefold. 
First, we introduce DFA, a low-cost and high-power method for auditing LLM relays through black-box access to suspect and trusted reference endpoints, without requiring model-owner cooperation.
Second, we develop a distributional fingerprinting test that aggregates probe-specific evidence into an approximate endpoint-level $p$-value, providing a statistical basis for audit decisions.
We further formulate likelihood-based model attribution under full substitution and joint estimation of substitute identity and substitution ratio under mixed routing.
Third, experiments on 16 LLMs demonstrate high full-substitution detection rates and low audit costs, and characterize how configuration overrides affect endpoint-consistency decisions. Simulations based on stored empirical fingerprints further demonstrate accurate attribution under full substitution and increasingly precise substitution-ratio estimates under constant-ratio mixed routing.

The remainder of this paper is organized as follows. \secref{sec:related_work} reviews related work on LLM fingerprinting and API auditing. \secref{sec:prelim} formalizes the auditing problem and specifies the threat model. \secref{sec:method} presents DFA and the likelihood-based procedures for model attribution and substitution-ratio estimation. \secref{sec:exp} evaluates detection performance, audit cost, robustness to configuration overrides, model attribution, and substitution-ratio estimation. \secref{sec:conclu} concludes the paper.

\section{Related Work}\label{sec:related_work}

\paragraph{LLM Fingerprinting.}
LLM fingerprinting seeks to identify a model or verify its provenance using model-specific signals in its outputs. A common distinction is between \emph{active fingerprinting} and \emph{passive fingerprinting}.
Active fingerprinting implants a verifiable mark into an LLM in advance, including injecting an instruction-following backdoor via fine-tuning~\citep{xu2024instructional}, embedding statistical watermarks into the decoding process~\citep{kirchenbauer2023watermark,ren2024robust}, and constructing false-claim-resistant ownership fingerprints~\citep{shao2026fit}.
However, these methods rely on the cooperation of the model owner and therefore cannot be deployed retrospectively by a third-party auditor.

Passive fingerprinting, by contrast, exploits behavioral regularities inherent in a deployed model without modifying the served model~\citep{shao2025sok}.
This line of work builds on the observation that LLM-generated text exposes strong “idiosyncrasies” signals—distributional quirks that enable authorship attribution and demonstrate that black-box outputs contain model identity information~\citep{sun2025idiosyncrasies}.
Existing methods differ primarily in the evidence extracted from API outputs.
\textsc{LLMmap}~\citep{pasquini2025llmmap} hand-crafts eight diagnostic prompts to elicit distinctive outputs from different LLMs, \textsc{TRAP}~\citep{gubri2024trap} repurposes adversarial suffixes as identity honeypots, and \textsc{ZeroPrint}~\citep{shao2025zeroprint} estimates local response sensitivity through semantically preserving prompt perturbations.
However, fingerprints constructed from long free-form completions, semantically structured perturbations, learned embeddings, or conspicuous fixed probes may shift under system prompts, reasoning modes, decoding settings, and benign provider-side wrappers~\citep{kbf}.
Our method constructs semantically simple finite-choice probes, repeatedly queries each probe, and treats the resulting response distributions as statistical evidence for endpoint-level consistency testing.
This formulation reduces reliance on rich response semantics while allowing robustness to deployment configurations to be evaluated explicitly in \secref{sec:exp_robustness}.

\paragraph{Auditing LLM APIs.}
A growing line of work investigates whether a black-box API faithfully serves the advertised model or service.
Earlier auditing methods detect behavioral drift across APIs through benchmark performance metrics such as task-level outcomes, confusion-matrix shifts, and performance changelists~\citep{chen2022modelchange,chen2023chatgpt,eyuboglu2024changelists}.
However, these aggregate performance metrics neither identify the served backend nor quantify partial routing.
\citep{met} formalizes this problem as \emph{model equality testing} (MET) and applies a string-kernel maximum mean discrepancy test to textual completions sampled from a suspect API and a reference distribution.
Nevertheless, \citep{tee} study substitution attacks involving quantization, smaller models, and version rollback, and show that simple output comparisons may fail under benign nondeterminism because observed behavior is jointly shaped by the underlying model, requested deployment configuration, and provider serving stack.
RUT~\citep{zhu2026rut} detects quantized substitutes using a rank-based uniformity test, but assumes local access to the reference model and its logits.
Concurrent with our work, KBF~\citep{kbf} audits black-box endpoints using self-calibrated knowledge-boundary probes and further considers constant-probability mixed routing.
However, its audit decision is based on a task-specific aggregation rule and does not provide a unified statistical interpretation for discrepancies distributed unevenly across probes.
Our method complements these auditing approaches in two respects: it requires only black-box query access to the official and relay endpoints, and it produces a statistically interpretable endpoint-level decision by combining probe-specific distributional evidence.

\section{Preliminaries}\label{sec:prelim}

\subsection{Problem: Endpoint Consistency Test}\label{sec:prelim_problem}

\begin{figure}[!htb]
  \centering
  \includegraphics[width=0.8\linewidth]{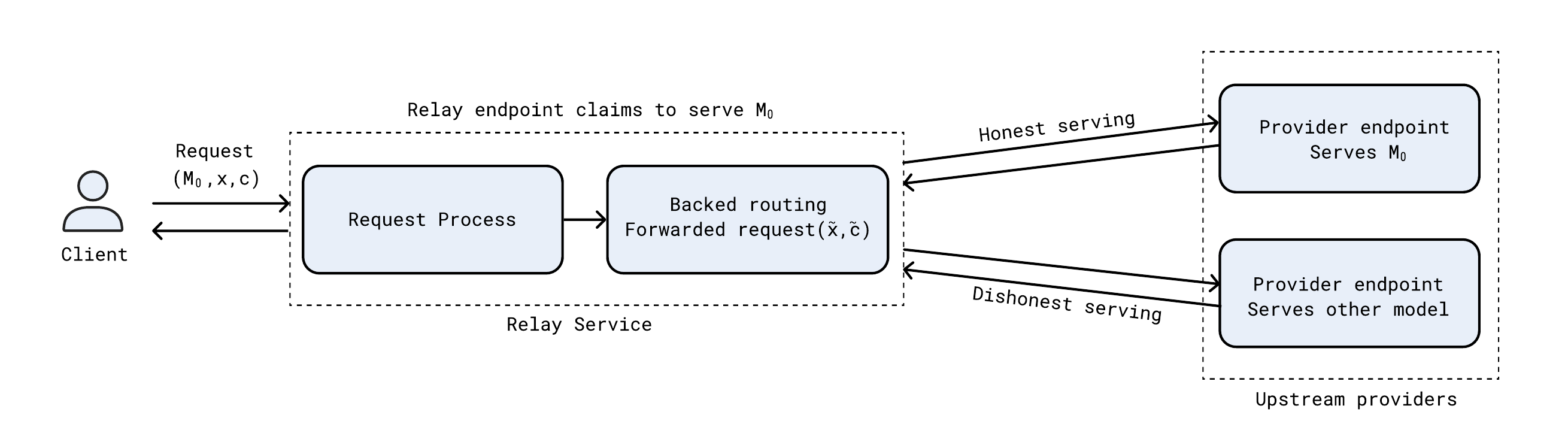}
  \caption{Request processing and backend routing in an LLM relay service. The client requests the advertised model $M_0$ with prompt $x$ and deployment configuration $c$. The relay may modify the request to $(\widetilde{x},\widetilde{c})$ and route it to an upstream endpoint serving $M_0$ or a substitute model. These internal processing and routing decisions are not directly observable to the client.}
  \label{fig:relay_system}
\end{figure}

We study whether an LLM relay operator actually serves its advertised model. We formalize this auditing problem as a \textit{black-box endpoint consistency test} between a suspect relay endpoint and an official reference endpoint~\citep{kbf}. Let $M_0$ denote the advertised model, $E_{\mathrm{ref}}$ the official reference endpoint trusted to serve $M_0$, and $E_{\mathrm{sus}}$ a suspect relay endpoint claiming to serve $M_0$.
The auditor queries each endpoint with a prompt $x$ and a user-specified deployment configuration $c\in\mathcal C$, such as a system prompt and decoding parameters. Given a request $(x,c)$, endpoint $E_e$ returns a textual response drawn from $P_e(\cdot\mid x,c)$, where $e\in\{\mathrm{ref},\mathrm{sus}\}$. \figref{fig:relay_system} illustrates the request-processing and backend-routing workflow of the relay service.

Repeated requests to the same endpoint with an identical $(x,c)$ may produce different outputs due to stochastic decoding and nondeterminism in the provider's serving stack. 
Let $\mathcal Q$ denote a prespecified set of audit requests. We compare the conditional response distributions $P_{\mathrm{ref}}(\cdot\mid x,c)$ and $P_{\mathrm{sus}}(\cdot\mid x,c)$ for requests in $\mathcal Q$ to assess whether the suspect endpoint's response behavior is consistent with the trusted reference endpoint.

\paragraph{Hypotheses.}
We formulate the endpoint consistency test through the following hypotheses:
\begin{equation}
  \label{eq:endpoint_equality}
  \begin{aligned}
    H_0 &:
    P_{\mathrm{sus}}(\cdot\mid x,c)
    = P_{\mathrm{ref}}(\cdot\mid x,c),
    \quad \forall\,(x,c)\in\mathcal Q, \\
    H_1 &:
    P_{\mathrm{sus}}(\cdot\mid x,c)
    \neq P_{\mathrm{ref}}(\cdot\mid x,c),
    \quad \exists\,(x,c)\in\mathcal Q.
  \end{aligned}
\end{equation}
$H_0$ states that the two endpoints induce the same conditional response distribution for every request in $\mathcal Q$, whereas $H_1$ holds when their distributions differ for at least one such request. 
These deviations may arise from model substitution or configuration overrides, as detailed in \secref{sec:prelim_threat_model}. We examine two forms of model substitution that may give rise to $H_1$.

\paragraph{Full substitution and model attribution.}
Under full substitution, the suspect endpoint serves a single substitute model $M_{\mathrm{sub}}$ for every audited request. This includes cross-family replacement and same-family downgrading.
When $H_0$ is rejected under the full-substitution assumption, we additionally examine the \emph{model attribution} objective: given a pool of candidate substitute models, the auditor identifies the model whose behavior best fits the observed responses from the suspect endpoint.

\paragraph{Mixed-routing substitution and substitution-ratio estimation.}
An adaptive relay may attempt to evade detection through mixed routing. Specifically, it serves $M_0$ for only a subset of requests and routes the remaining requests to a substitute model $M_{\mathrm{sub}}$.  When routing preserves each selected model's reference response distribution, the relay's conditional response distribution is
\begin{equation}
  P_{\mathrm{sus}}(\cdot\mid x,c) =
  \bigl[1-\pi(x,c)\bigr]P_{\mathrm{ref}}(\cdot\mid x,c) + \pi(x,c)P_{\mathrm{sub}}(\cdot\mid x,c),
  \label{eq:mixed_routing_distribution}
\end{equation}
where $P_{\mathrm{sub}}(\cdot\mid x,c)$ is the substitute model's response distribution, and $\pi(x,c)\in[0,1]$ denotes the conditional probability that request $(x,c)$ is routed to $M_{\mathrm{sub}}$. Routing decisions may depend on the prompt, service tier, system load, or the perceived likelihood that a request is part of an audit~\citep{kbf}. Mixed routing can make substitution harder to detect because the substitute model contributes only a fraction of the observed responses~\citep{met,kbf}. For substitution-ratio estimation, we consider the routing probability independent of the request, i.e., $\pi(x,c)\equiv\pi$. When $H_0$ is rejected under this model, the auditor seeks to estimate $\pi$ and, when the substitute identity is unknown, identify the candidate model that best explains the observed responses.

\subsection{Threat Model}\label{sec:prelim_threat_model}

\begin{figure}[!htb]
  \centering
  \includegraphics[width=0.8\linewidth]{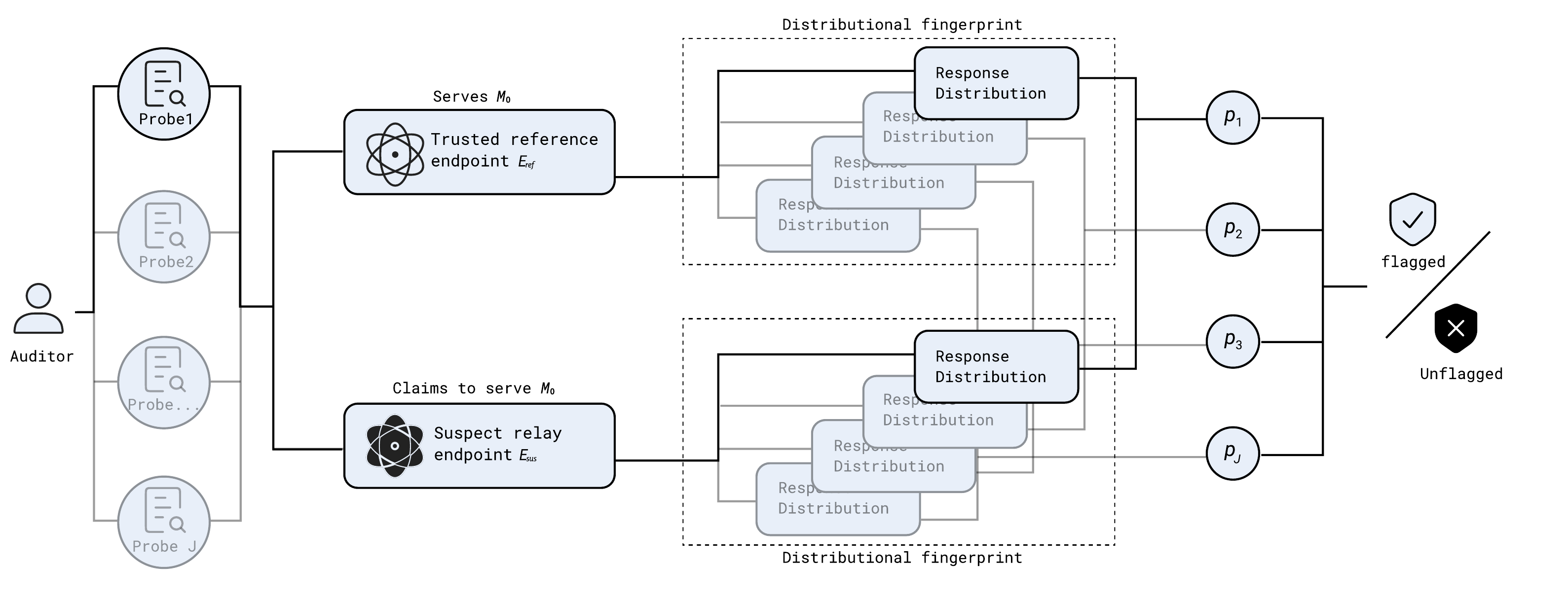}
  \caption{Overview of the distributional fingerprinting audit. The auditor repeatedly queries a trusted reference endpoint and a suspect relay endpoint using the same audit probes to estimate their response distributions. Probe-specific tests compare these distributions and produce $p$-values, which are combined into an endpoint-level audit decision. Flagged and unflagged indicate rejection and non-rejection of endpoint consistency, respectively.}
  \label{fig:threat_model}
\end{figure}

We consider an auditing setting involving an auditor and a potentially adversarial relay operator, as shown in \figref{fig:threat_model}. The relay operator provides query access to the suspect endpoint $E_{\mathrm{sus}}$. The auditor assesses the relay's claim by comparing the suspect endpoint's responses with those from the trusted reference endpoint $E_{\mathrm{ref}}$.

\paragraph{Auditor capabilities.}
The auditor has \emph{black-box} query access to the suspect endpoint: it can submit requests $(x,c)$ and observe the endpoint's responses, but has no access to model weights, per-token probabilities, or other internal execution details. This setting represents an independent third-party auditor who receives no cooperation from the relay operator and interacts with the suspect endpoint through its public API. 
We assume that the auditor also has black-box query access to the reference endpoint $E_{\mathrm{ref}}$, trusted to serve the advertised model $M_0$, and a set of trusted candidate endpoints $\mathcal R_{\mathrm{cand}}=\{E_{\mathrm{cand}}^k\}_{k=1}^{K}$. Each $E_{\mathrm{cand}}^k$ serves a candidate substitute model $M_k$.

\paragraph{Adaptive relay.}
The suspect relay controls the backend serving logic of $E_{\mathrm{sus}}$ upon receiving a user request $(x,c)$. It may inspect the incoming prompt $x$ to identify potential audit probes and selectively route these requests to the advertised model, thereby reducing the observable evidence of inconsistency and potentially evading detection~\citep{kbf}. Such selective routing can make the substitution ratio on audited traffic differ from that on ordinary user traffic. For the requested configuration $c$, the relay may override system prompts and decoding parameters. It may further alter the serving workflow by inserting hidden prompts or wrappers, augmenting generation with retrieval or tools, or transforming returned responses. We assess the robustness of our audit under configuration overrides in \secref{sec:exp_robustness}.

\paragraph{Auditing scope.}
Our audit assesses endpoint behavior rather than verifying the underlying model weights. 
If the reference endpoint incorporates retrieval or tools, their effects are part of the reference behavior. 
Conversely, if the suspect relay adds an undisclosed wrapper, retrieval component, or tool to a claimed bare-model endpoint, this can alter the suspect endpoint's behavior even when the user-requested configuration remains unchanged.
In this purely black-box setting, the endpoint consistency test alone cannot determine whether an observed discrepancy arises from model substitution, retrieval, tool use, or other serving-stack changes.

\section{Methodology: Distributional Fingerprinting Audit}\label{sec:method}

\subsection{Endpoint-level Distributional Fingerprint}\label{sec:method_intuition}

\paragraph{Intuition.}
Our method is motivated by the observation that LLM outputs exhibit model-specific behavioral patterns~\citep{sun2025idiosyncrasies}. These patterns also emerge in responses to semantically simple choice prompts. For example, when asked to select a number at random from a given set, LLMs tend to favor certain numbers, with preferences varying across models~\citep{renda2023sampling,coronado2025deterministic}. 
Such systematic differences motivate the use of response distributions as behavioral fingerprints for model identification. 
To capture these differences, we carefully craft a suite of \emph{audit probes}, each consisting of a semantically simple finite-choice prompt and a specified deployment configuration.
We repeatedly query each endpoint with these probes and estimate a response distribution for each probe from the collected responses.
The resulting collection of distributions forms an endpoint-level fingerprint, which we compare statistically with that of the claimed model's trusted reference endpoint to assess distributional consistency.

\paragraph{Constructing distributional fingerprints.}
We craft an audit probe suite $\mathcal Q=\{(x_j,c_j)\}_{j=1}^{J}$, where prompt $x_j$ asks the endpoint to select from a finite set of options, such as letters, numbers, or colors, and $c_j\in\mathcal C$ specifies the requested deployment configuration.
For the $j$th probe, a prespecified parser $g_j$ maps every response to exactly one category in the finite set $\mathcal Z_j$. Responses that cannot be parsed or violate the required format are assigned to an invalid category within $\mathcal Z_j$, preserving instruction-following failures as part of the endpoint's observable behavior.
We use the same probe suite, category sets, and parsers for all endpoints to construct comparable fingerprints.

For an endpoint $E_e$, let $y_{e,j}^{(i)}$ denote its textual response to the $i$th request using probe $(x_j,c_j)$, and let $z_{e,j}^{(i)}=g_j(y_{e,j}^{(i)})\in\mathcal Z_j$ denote the corresponding parsed category. 
For each fixed endpoint--probe pair, we model the parsed responses as independent draws from a common categorical distribution:
\begin{equation}
  z_{e,j}^{(i)}
  \overset{\mathrm{i.i.d.}}{\sim}
  \operatorname{Cat}(\boldsymbol{\theta}_{e,j}), %
\end{equation}
where $\boldsymbol{\theta}_{e,j}=(\theta_{e,j}(z))_{z\in\mathcal Z_j}$ is the vector of category probabilities of endpoint $E_e$ on probe $(x_j,c_j)$, with each component $\theta_{e,j}(z)=P_e(\{y:g_j(y)=z\}\mid x_j,c_j)$ giving the probability that the endpoint's textual response is parsed as category $z$.
We define the distributional fingerprint of $E_e$ as the collection of its categorical response distributions over $\mathcal{Q}$, that is, $\mathrm{FP}(E_e)=\{\operatorname{Cat}(\boldsymbol{\theta}_{e,j})\}_{j=1}^{J}$. 
Under $H_0$, the reference and suspect endpoints induce the same categorical distribution for every probe and therefore share the same distributional fingerprint.

In practice, the auditor submits $N_{e,j}$ repeated requests to endpoint $E_e$ using each probe $(x_j,c_j)$ and estimates the categorical distribution from the observed frequencies of $\{z_{e,j}^{(i)}\}_{i=1}^{N_{e,j}}$. 
The resulting estimated distributions form an empirical fingerprint, whose construction procedure is detailed in Algorithm~\ref{alg:fingerprint_construction}. Before auditing a suspect relay, we construct and register empirical fingerprints for the advertised model and candidate substitutes using their trusted endpoints.

\begin{algorithm}[t]
\caption{Empirical Distributional Fingerprint Construction}
\label{alg:fingerprint_construction}
\DontPrintSemicolon
\KwOracle{Black-box endpoint $E_e\in\{E_{\mathrm{sus}},E_{\mathrm{ref}}\}\cup\mathcal R_{\mathrm{cand}}$}
\KwIn{Fixed audit suite $\mathcal Q=\{(x_j,c_j)\}_{j=1}^{J}$; category sets $\{\mathcal Z_j\}_{j=1}^{J}$, each including an invalid category; prespecified parsers $\{g_j\}_{j=1}^{J}$; request counts $\{N_{e,j}\geq 1\}_{j=1}^{J}$}
\KwOut{Empirical fingerprint record $\mathrm{FP}(E_e)$ containing probe specifications, raw counts, sample sizes, and empirical distributions}

Initialize an empty fingerprint record $\mathrm{FP}(E_e)$\;
\For{$j=1,\ldots,J$}{
    $n_{e,j}(z)\leftarrow 0$ for all $z\in\mathcal Z_j$\;
    \For{$i=1,\ldots,N_{e,j}$}{
        Query $E_e$ with probe $(x_j,c_j)$ to obtain response $y$\;
        $z\leftarrow g_j(y)$\;
        $n_{e,j}(z)\leftarrow n_{e,j}(z)+1$\;
    }
    $\widehat{\theta}_{e,j}(z)\leftarrow n_{e,j}(z)/N_{e,j}$ for every $z\in\mathcal Z_j$\;
    Store $\bigl((x_j,c_j),\mathcal Z_j,g_j,N_{e,j},n_{e,j},\widehat{\boldsymbol{\theta}}_{e,j}\bigr)$ as the record for probe $j$ in $\mathrm{FP}(E_e)$\;
}
\KwRet{$\mathrm{FP}(E_e)$}\;
\end{algorithm}

\subsection{Distributional Testing for Model Substitution}\label{sec:method_test}

Model substitution may alter the response distribution of one or more probes. We test for these changes by comparing the reference and suspect fingerprints probe by probe, then combining the resulting probe-specific evidence into an endpoint-level audit decision.

\paragraph{Probe-specific evidence construction.}
For each probe $(x_j,c_j)$, we test whether the reference and suspect endpoints induce the same categorical response distribution under the requested configuration $c_j$:
\begin{equation}
  H_{0,j}: \boldsymbol{\theta}_{\mathrm{ref},j}=\boldsymbol{\theta}_{\mathrm{sus},j}, \qquad
  H_{1,j}: \boldsymbol{\theta}_{\mathrm{ref},j}\neq \boldsymbol{\theta}_{\mathrm{sus},j}.
\end{equation}
We summarize the $N_{\mathrm{ref},j}$ reference responses and $N_{\mathrm{sus},j}$ suspect responses in a two-row contingency table:
\begin{equation}
  \label{eq:probe_contingency}
  \mathbf O_j=
  \begin{bmatrix}
    n_{\mathrm{ref},j}(z_1) & \cdots & n_{\mathrm{ref},j}(z_{C_j})\\
    n_{\mathrm{sus},j}(z_1) & \cdots & n_{\mathrm{sus},j}(z_{C_j})
  \end{bmatrix},
\end{equation}
where $n_{e,j}(z)=\sum_{i=1}^{N_{e,j}}\mathbb I[z_{e,j}^{(i)}=z]$ is the number of times category $z$ is observed for endpoint $E_e$, with $\sum_{z\in\mathcal Z_j}n_{e,j}(z)=N_{e,j}$. The table retains the $C_j$ categories observed at least once across the two endpoints and omits categories with zero counts in both rows.

Under $H_{0,j}$, both rows share a common categorical distribution. We estimate its category probabilities by the pooled proportions $\widehat{\theta}_j(z_r)=[n_{\mathrm{ref},j}(z_r)+n_{\mathrm{sus},j}(z_r)]/(N_{\mathrm{ref},j}+N_{\mathrm{sus},j})$, giving the fitted expected counts $\mu_{e,j}(z_r)=N_{e,j}\widehat{\theta}_j(z_r)$.
We quantify the discrepancy between the observed and fitted counts using Pearson's chi-squared statistic~\citep{pearson1900criterion}:
\begin{equation}
  \label{eq:pearson_statistic}
  \chi_j^2 = \sum_{e\in\{\mathrm{ref},\mathrm{sus}\}} \sum_{r=1}^{C_j}
  \frac{\bigl(n_{e,j}(z_r)-\mu_{e,j}(z_r)\bigr)^2}{\mu_{e,j}(z_r)}.
\end{equation}
For $C_j>1$ and sufficiently large expected cell counts, the statistic is asymptotically distributed as $\chi^2_{C_j-1}$ under $H_{0,j}$, that is, $\chi_j^2 \overset{\mathrm{a}}{\sim} \chi^2_{C_j-1}$~\citep{pearson1900criterion}.
We compute the corresponding probe-specific $p$-value from the upper-tail probability:
\begin{equation}
  \label{eq:probe_pvalue}
  p_j=\Pr[\chi^2_{C_j-1}\geq\chi_j^2].
\end{equation}
A small $p_j$ provides evidence of a distributional discrepancy between the reference and suspect endpoints on the $j$th probe.

\paragraph{Evidence aggregation across probes.}
Probes may differ substantially in their ability to distinguish a given pair of suspect and reference endpoints~\citep{met}. An adaptive relay may recognize certain audit probes and route them to the advertised model, masking substitution on those probes. Other probes may elicit similar response distributions across models and therefore have little discriminative power. Consequently, evidence of substitution may be concentrated in only a subset of the audit suite. We combine the probe-specific $p$-values across the audit suite into an endpoint-level evidence statistic using the Cauchy combination test (CCT)~\citep{liu2019acat,cct}:
\begin{equation}
  \label{eq:cauchy_statistic}
  T_{\mathrm{Cauchy}}=\frac{1}{J}\sum_{j=1}^{J}\tan\!\left[\pi\left(\frac{1}{2}-p_j\right)\right],
\end{equation}
where the Cauchy transformation $\tan[\pi(0.5-p_j)]$ maps a small $p_j$ to a large positive value, whereas a $p_j$ near $1/2$ contributes close to zero. 
This nonlinear transformation allows discriminative probes to dominate the endpoint-level statistic, thereby mitigating the dilution of evidence by less informative probes.

In practice, the probe-specific $p$-values may be dependent because requests can share routing states or serving conditions. The CCT is well suited to this setting because, under $H_0$ and its regularity conditions, the upper tail of $T_{\mathrm{Cauchy}}$ can be approximated by a standard Cauchy distribution~\citep{cct}. We use this approximation to compute the combined $p$-value:
\begin{equation}
  \label{eq:cauchy_pvalue}
  p_{\mathrm{Cauchy}}=\frac{1}{2}-\frac{1}{\pi}\arctan\!\left(T_{\mathrm{Cauchy}}\right),
\end{equation}
where $\arctan(\cdot)$ denotes the principal inverse tangent function.
A smaller $p_{\mathrm{Cauchy}}$ indicates stronger evidence of a distributional discrepancy between the suspect and reference endpoints across the audited probes. We reject endpoint consistency when $p_{\mathrm{Cauchy}}\leq\alpha$, where $\alpha$ is a prespecified nominal significance level specifying the target probability of falsely rejecting distributional consistency under $H_0$.
We list the audit procedure in Algorithm~\ref{alg:distributional_audit}, with implementation details provided in Appendix~B.6.

\begin{algorithm}[t]
\caption{Distributional Fingerprint Audit}
\label{alg:distributional_audit}
\DontPrintSemicolon
\KwOracle{Suspect relay endpoint $E_{\mathrm{sus}}$}
\KwIn{Stored reference fingerprint $\mathrm{FP}(E_{\mathrm{ref}})$; request counts $\{N_{\mathrm{sus},j}\geq 1\}_{j=1}^{J}$; nominal significance level $\alpha\in(0,1)$; lower $p$-value threshold $\epsilon=0.0001$}
\KwOut{Combined $p$-value $p_{\mathrm{Cauchy}}$ and audit decision $\delta_\alpha$ ($1$: reject $H_0$; $0$: do not reject)}
\textbf{Stage I: Suspect fingerprint collection}\;
Retrieve $\mathcal Q$, $\{\mathcal Z_j\}$, and $\{g_j\}$ from $\mathrm{FP}(E_{\mathrm{ref}})$\;
Construct $\mathrm{FP}(E_{\mathrm{sus}})$ using Algorithm~\ref{alg:fingerprint_construction} with these specifications and request counts $\{N_{\mathrm{sus},j}\}$\;
\textbf{Stage II: Probe-specific distribution comparison}\;
\For{$j=1,\ldots,J$}{
    Retrieve counts $n_{e,j}$ and sample sizes $N_{e,j}$ for probe $j$ from both fingerprints\;
    Retain the categories observed at least once across the two endpoints in $\mathcal Z_j^+$\;
    $C_j\leftarrow|\mathcal Z_j^+|$\;
    \eIf{$C_j=1$}{
        $p_j\leftarrow 1$\;
    }{
        Form the count table $\mathbf O_j$ over $\mathcal Z_j^+$ using Eq.~\ref{eq:probe_contingency}\;
        \ForEach{$z\in\mathcal Z_j^{+}$}{
            $\widehat\theta_j(z)\leftarrow [n_{\mathrm{ref},j}(z)+n_{\mathrm{sus},j}(z)]/(N_{\mathrm{ref},j}+N_{\mathrm{sus},j})$\;
            $\mu_{e,j}(z)\leftarrow N_{e,j}\widehat\theta_j(z)$ for $e\in\{\mathrm{ref},\mathrm{sus}\}$\;
        }
        Compute Pearson's statistic $\chi_j^2$ using Eq.~\ref{eq:pearson_statistic}\;
        Compute $p_j$ using Eq.~\ref{eq:probe_pvalue}\;
    }
}
\textbf{Stage III: Evidence aggregation and decision}\;
$\widetilde p_j\leftarrow\max\{\epsilon,p_j\}$ for $j=1,\ldots,J$\;
Compute $T_{\mathrm{Cauchy}}$ using Eq.~\ref{eq:cauchy_statistic} with $\widetilde p_j$ in place of $p_j$\;
Compute $p_{\mathrm{Cauchy}}$ using Eq.~\ref{eq:cauchy_pvalue}\;
$\delta_\alpha\leftarrow\mathbb I[p_{\mathrm{Cauchy}}\leq\alpha]$\;
\KwRet{$(p_{\mathrm{Cauchy}},\delta_\alpha)$}\;
\end{algorithm}

\subsection{Likelihood-based Model Attribution and Substitution-Ratio Estimation}\label{sec:method_inference}

After rejecting endpoint consistency, the auditor further seeks to reveal the source of the behavioral discrepancy.
Specifically, the auditor aims to identify the substitute model and quantify its contribution to the audited traffic.
We address these objectives through a likelihood-based framework using the registered fingerprints: full substitution corresponds to selecting a single candidate, while mixed routing requires estimating both the candidate identity and its substitution ratio.

For likelihood-based inference, we index the estimated category probabilities by model: $\widehat{\boldsymbol{\theta}}_{0,j}$ denotes the estimate obtained from $E_{\mathrm{ref}}$, while $\widehat{\boldsymbol{\theta}}_{k,j}$ denotes the estimate obtained from $E_{\mathrm{cand}}^k$ for $k\in\{1,\ldots,K\}$.
In the experiments, the probabilities used for likelihood evaluation are smoothed as described in \secref{sec:exp_setup}.
For a candidate substitute $M_s$ and substitution ratio $\pi$, these fingerprints specify the response probabilities expected from a relay that independently routes each request to $M_0$ with probability $1-\pi$ and to $M_s$ with probability $\pi$:
\begin{equation}
  \theta_{s,j}^{\mathrm{mix}}(z;\pi)
  =(1-\pi)\widehat{\theta}_{0,j}(z)
  +\pi\widehat{\theta}_{s,j}(z).
\end{equation}
Here, $\pi=0$ corresponds to honest serving, $\pi=1$ to full substitution, and $0<\pi<1$ to mixed routing.
Given the suspect endpoint's response counts, we define the log-likelihood under independent observations as
\begin{equation}
  \label{eq:likelihood}
  \ell(s,\pi)
  =\sum_{j=1}^{J}\sum_{z\in\mathcal Z_j}
  n_{\mathrm{sus},j}(z)\log\theta_{s,j}^{\mathrm{mix}}(z;\pi),
\end{equation}
where terms independent of $(s,\pi)$ are omitted. A larger log-likelihood indicates that the candidate model and substitution ratio better explain the suspect endpoint's observed responses. By combining observations across the audit suite, this criterion provides a common basis for model attribution and substitution-ratio estimation.

\paragraph{Model attribution.}
Under full substitution, the relay serves a single substitute model for all requests, corresponding to $\pi=1$. We attribute its behavior to the candidate with the largest log-likelihood:
\begin{equation}
  \label{eq:full_substitution_attribution}
  \widehat{s}
  =\arg\max_{s\in\{1,\ldots,K\}}\ell(s,1).
\end{equation}
The resulting $\widehat{s}$ identifies the candidate whose reference fingerprint best explains the suspect responses.

\paragraph{Substitution-ratio estimation.}
Under constant-ratio mixed routing, we jointly estimate the substitute identity and substitution ratio:
\begin{equation}
  \label{eq:joint_mixture_estimation}
  (\widehat{s},\widehat{\pi})
  =\arg\max_{\substack{s\in\{1,\ldots,K\},\ \pi\in[0,1]}}
  \ell(s,\pi).
\end{equation}
For each candidate, we first find the substitution ratio that maximizes its log-likelihood and then select the candidate with the largest maximized log-likelihood. The resulting $\widehat{\pi}$ estimates the fraction of audited requests served by the substitute.
In practice, a relay may declare that requests are routed to a specified fallback model $M_{\mathrm{sub}}$ during overload. In this case, the candidate index $s$ is fixed, and~\eqref{eq:joint_mixture_estimation} reduces to $\widehat{\pi}=\arg\max_{\pi\in[0,1]}\ell(s,\pi)$. This estimate quantifies the use of the declared fallback model over the audited traffic, providing evidence for auditing the relay's fallback policy. Algorithm~\ref{alg:likelihood_inference} summarizes the two inference procedures, and an extension to multiple substitute models is provided in Appendix~B.4.
\begin{algorithm}[t]
\caption{Likelihood-based Model Attribution and Substitution-Ratio Estimation}
\label{alg:likelihood_inference}
\DontPrintSemicolon
\KwIn{Registered empirical fingerprints $\mathrm{FP}(E_{\mathrm{ref}})$ and $\{\mathrm{FP}(E_{\mathrm{cand}}^k)\}_{k=1}^{K}$; suspect response counts $\{n_{\mathrm{sus},j}\}_{j=1}^{J}$; candidate index set $\mathcal S=\{1,\ldots,K\}$, or $\mathcal S=\{s_0\}$ for a declared fallback model $M_{s_0}$; inference setting: full substitution or mixed routing}
\KwOut{Substitute index $\widehat{s}$ ($\varnothing$ if undetermined) and substitution ratio $\widehat{\pi}$}

Form $\ell(s,\pi)$ from the registered empirical fingerprints and suspect counts using Eq.~\ref{eq:likelihood}\;

\ForEach{$s\in\mathcal S$}{
    \eIf{the setting is full substitution}{
        $\widehat{\pi}_s\leftarrow 1$\;
    }{
        $\widehat{\pi}_s\leftarrow\displaystyle\arg\max_{\pi\in[0,1]}\ell(s,\pi)$\;
    }
    $L_s\leftarrow\ell(s,\widehat{\pi}_s)$\;
}

$\widehat{s}\leftarrow\displaystyle\arg\max_{s\in\mathcal S}L_s$\;
$\widehat{\pi}\leftarrow\widehat{\pi}_{\widehat{s}}$\;

\If{$\widehat{\pi}=0$ and $|\mathcal S|>1$}{
    $\widehat{s}\leftarrow\varnothing$\;
}
\KwRet{$(\widehat{s},\widehat{\pi})$}\;
\end{algorithm}

\section{Experiments}
\label{sec:exp}

We evaluate \method through five research questions (RQs):
\begin{enumerate}[leftmargin=2.8em, labelwidth=2.4em, labelsep=0.4em, align=left, itemsep=0.3em, topsep=0.3em, parsep=0pt, label=\textbf{RQ\arabic*:}, ref=RQ\arabic*]
  \item \label{rq:1} \textbf{Detection performance.}
        Can \method detect full model substitution, including substitution within the same model family?
  \item \label{rq:2} \textbf{Cost analysis.}
        How does detection performance vary with the request budget, and what are the costs of reference collection and individual audits?
  \item \label{rq:3} \textbf{Robustness.}
        How do deployment-configuration overrides affect the non-rejection rate of endpoint consistency?
  \item \label{rq:4} \textbf{Model attribution.}
        Can \method identify the substitute model among the reference candidates?
  \item \label{rq:5} \textbf{Substitution-ratio estimation.}
        Can \method estimate the proportion of requests routed to a substitute model under mixed routing?
\end{enumerate}

\subsection{Experimental Setup}
\label{sec:exp_setup}

We evaluate \method in a controlled setting with fixed endpoints, provider routes, audit probes, and a prespecified decision rule.
This setting provides a common basis for assessing full substitution, audit cost, configuration overrides, model attribution, and mixed routing.

\paragraph{Models.}

The primary benchmark includes 16 LLMs from 10 model families, covering proprietary and open-weight models of different sizes and price levels.
Multiple models from the \texttt{Claude}, \texttt{DeepSeek}, \texttt{GLM}, \texttt{GPT}, and \texttt{MiMo} families allow us to examine both cross-family and within-family substitution.
\Cref{tab:model_inventory} lists the evaluated models, their official providers, and published API prices.

\begin{longtblr}[
  caption = {Overview of the 16 LLMs from 10 model families used in the primary benchmark. For each model, we report its family, releasing organization, and published input and output API prices in USD per million tokens.},
  label = {tab:model_inventory},
  note{*} = {Prices were checked on 18 September 2026 and apply to uncached text input and text output at the lowest input-length tier.},
] {
  width = \textwidth,
  colspec = {X[3,halign=l,valign=m] X[0.9,halign=l,valign=m] X[1.2,halign=l,valign=m] X[0.8,halign=r,valign=m] X[0.8,halign=r,valign=m]},
  colsep = 4pt,
  rowhead = 2,
  rows = {font=\small},
  row{1,2} = {font=\small\bfseries},
}
\toprule
\SetCell[r=2]{l} Model & \SetCell[r=2]{l} Model Family & \SetCell[r=2]{l} Model Provider & \SetCell[c=2]{c} Price$^{*}$ (\$/M Tokens) & \\
\cmidrule{4-5}
 & & & Input & Output \\
\midrule
\texttt{claude-sonnet-5.5} & Claude & Anthropic & 2.00 & 10.00 \\
\texttt{claude-opus-5.5} & Claude & Anthropic & 4.00 & 20.00 \\
\texttt{deepseek-flash} & DeepSeek & DeepSeek & 0.30 & 1.20 \\
\texttt{deepseek-v4-pro} & DeepSeek & DeepSeek & 1.74 & 3.84 \\
\texttt{glm-5.2} & GLM & Z.ai & 1.40 & 4.40 \\
\texttt{glm-5.3} & GLM & Z.ai & 1.40 & 4.40 \\
\texttt{gpt-5.6-luna} & GPT & OpenAI & 0.20 & 1.20 \\
\texttt{gpt-6-luna} & GPT & OpenAI & 0.10 & 0.50 \\
\texttt{gpt-6-sol} & GPT & OpenAI & 2.00 & 10.00 \\
\texttt{gpt-oss-120b} & GPT-OSS & OpenAI & 0.03 & 0.15 \\
\texttt{muse-spark-1.3} & Muse Spark & Meta & 1.25 & 4.25 \\
\texttt{hy3} & Hunyuan & Tencent & 0.13 & 0.53 \\
\texttt{mimo-v2.5} & MiMo & Xiaomi & 0.14 & 0.28 \\
\texttt{mimo-v2.5-pro} & MiMo & Xiaomi & 0.44 & 0.87 \\
\texttt{minimax-m3} & MiniMax & MiniMax & 0.30 & 1.20 \\
\texttt{nemotron-3-ultra} & Nemotron & NVIDIA & 0.50 & 2.20 \\
\bottomrule
\end{longtblr}

\paragraph{Reference endpoints.}

We construct the reference fingerprint of each claimed model using Algorithm~\ref{alg:fingerprint_construction}.
Reference responses are collected from the corresponding official API, except for \texttt{GPT-OSS} and \texttt{Nemotron}, for which we use self-hosted endpoints.
\Cref{tab:model_inventory} identifies the organization releasing each model or checkpoint, and Appendix~A provides the hosting details.

\paragraph{Implementation details.}

Our audit probes use five finite-choice prompts: two in Chinese and three in English.
For each reference endpoint and audit probe, we collect 1,000 responses at temperature $1.0$ and $\texttt{top\_p} = 0.8$, with a maximum completion length of 1,024 tokens.
We disable thinking where supported and otherwise use the lowest supported reasoning-effort setting; provider-specific settings are detailed in Appendix~A.
A prespecified parser maps each response to a valid response category, with malformed, missing, unmatched, and out-of-range responses assigned to an explicit \emph{outlier} category.
Appendix~B.1 provides the complete probe texts, response categories, and regular-expression-based parsing procedure.

For~\ref{rq:4}--\ref{rq:5}, we generate new audit category counts in each simulation while keeping the stored empirical fingerprints fixed. 
To avoid zero probabilities in likelihood evaluation, we add $0.5$ to the stored count of every category, including the outlier category, and normalize the adjusted counts within each model–probe pair. 
This additive smoothing is used only for likelihood computation; simulated audit counts are generated from the original empirical frequencies.

\paragraph{Evaluation metrics.}

We evaluate substitution detection and likelihood-based inference separately.
For substitution detection, a false positive occurs when the auditor rejects endpoint consistency for an endpoint serving the claimed model under the reference configuration, whereas a true positive occurs when it rejects endpoint consistency under full substitution.
We report the corresponding empirical rates at a nominal significance level of $\alpha=0.05$.
For~\ref{rq:2}, we additionally report detection success, defined as the correct-decision rate averaged equally over the evaluated serving models.

For likelihood-based inference, we report top-1 attribution accuracy under full substitution and the root-mean-square error (RMSE) of the estimated substitution ratio under mixed routing.
Top-1 attribution accuracy measures the fraction of audits in which the selected candidate matches the actual substitute model.
RMSE measures the root-mean-square difference between estimated and true substitution ratios.
We give each ordered model pair equal weight and evaluate all simulated audits, including those for which the endpoint-consistency test fails to reject the null hypothesis.

\begin{figure}[thbp]
  \centering
  \includegraphics[width=0.9\linewidth]{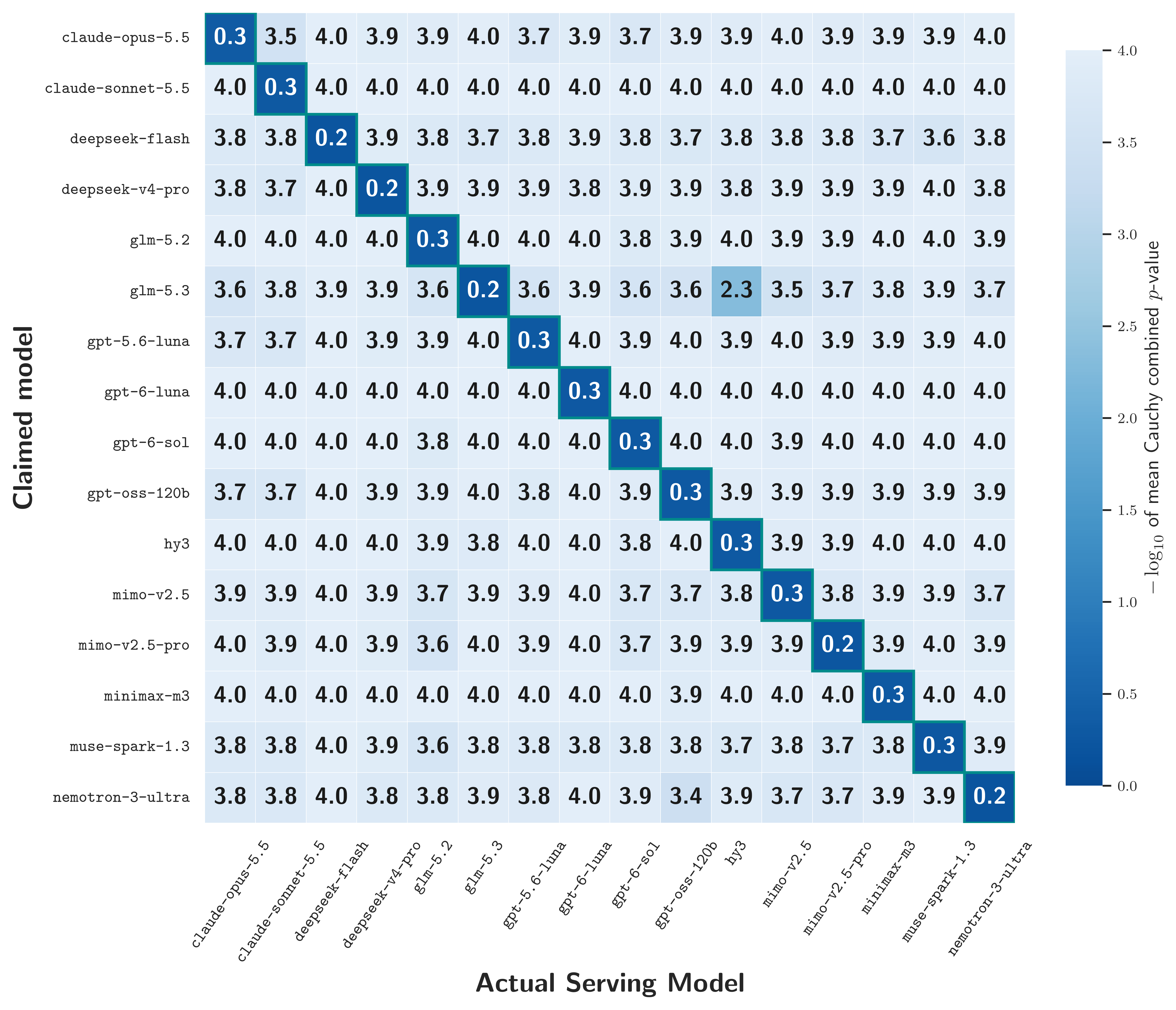}
  \caption{
    Pairwise audit evidence with 50 responses per probe.
    Each cell reports $-\log_{10}\overline p$, where $\overline p$ is the mean Cauchy-combined $p$-values over 1,000 audits.
    Rows indicate the claimed model; columns indicate the actual serving model.
    Outlined diagonal cells represent honest serving, and off-diagonal cells represent full substitution.
  }
  \label{fig:exp1_acceptance_matrix}
\end{figure}

\subsection{\ref{rq:1}: Detection Performance}
\label{sec:exp_substitution}

We examine whether \method distinguishes honest serving from full substitution with a fixed audit budget.
Each of the 16 models serves in turn as the claimed model $M_0$.
We compare its stored reference fingerprint with responses from a suspect endpoint serving either $M_0$, as a same-model control, or one of the remaining 15 models, as a full substitute.
This design covers 240 ordered substitution pairs.

Each audit collects 50 fresh API responses for each of the five probes, giving 250 requests in total, and computes the Cauchy-combined $p$-value using \Cref{alg:distributional_audit}.
We repeat the audit 1,000 times for each claimed–serving pair, and summarize the resulting evidence by $-\log_{10}\overline p$, where $\overline p$ is the mean Cauchy-combined $p$-value across audits.
Larger values correspond to smaller mean $p$-values and therefore stronger aggregate evidence against endpoint consistency.
Diagonal entries correspond to honest serving, and off-diagonal entries correspond to full substitution.

\Cref{fig:exp1_acceptance_matrix} shows a clear separation between same-model controls and full substitution.
Diagonal values are approximately $0.2$–$0.3$, whereas most off-diagonal values lie between $3$ and $4$, corresponding to mean $p$-values on the order of $10^{-3}$–$10^{-4}$.
The separation also appears within the model family, including the \texttt{Claude}, \texttt{DeepSeek}, and \texttt{MiMo} pairs, indicating that the distributional fingerprints capture differences among individual models of the same family.
Notably, the smallest off-diagonal value is $2.322$, obtained when \texttt{glm-5.3} is the claimed model and \texttt{hy3} supplies the responses.
Although this pair produces weaker evidence than most substitutions, its mean evidence remains separated from the same-model controls.
The pairwise empirical non-rejection rates in Appendix~B.2 characterize how these differences translate into audit decisions at $\alpha = 0.01$ and $0.05$.

\subsection{\ref{rq:2}: Cost Analysis}
\label{sec:exp_cost}

We examine the request budget needed for reliable audit decisions and separate the one-time cost of reference collection from the cost of subsequent audits.
With five probes and $n$ responses per probe, each audit requires $5n$ requests to the suspect endpoint.
Constructing each reference fingerprint requires 1,000 responses per probe, requiring 5,000 requests in total. The stored fingerprint is then reused across audits.

\paragraph{Detection success versus budget.}

We vary the number of responses per probe $n$ up to $1,000$ to examine how the request budget affects detection success.
For every claimed–serving pair and budget, we conduct $1,000$ audits with newly collected API responses.
For each claimed model, detection success is the correct-decision rate averaged equally over the 16 serving models: non-rejection is correct for the same-model control, and rejection is correct for each of the 15 substitutes.

\begin{figure}[htbp]
  \centering
  \begin{minipage}[c]{0.55\linewidth}
    \centering
    \includegraphics[width=\linewidth]{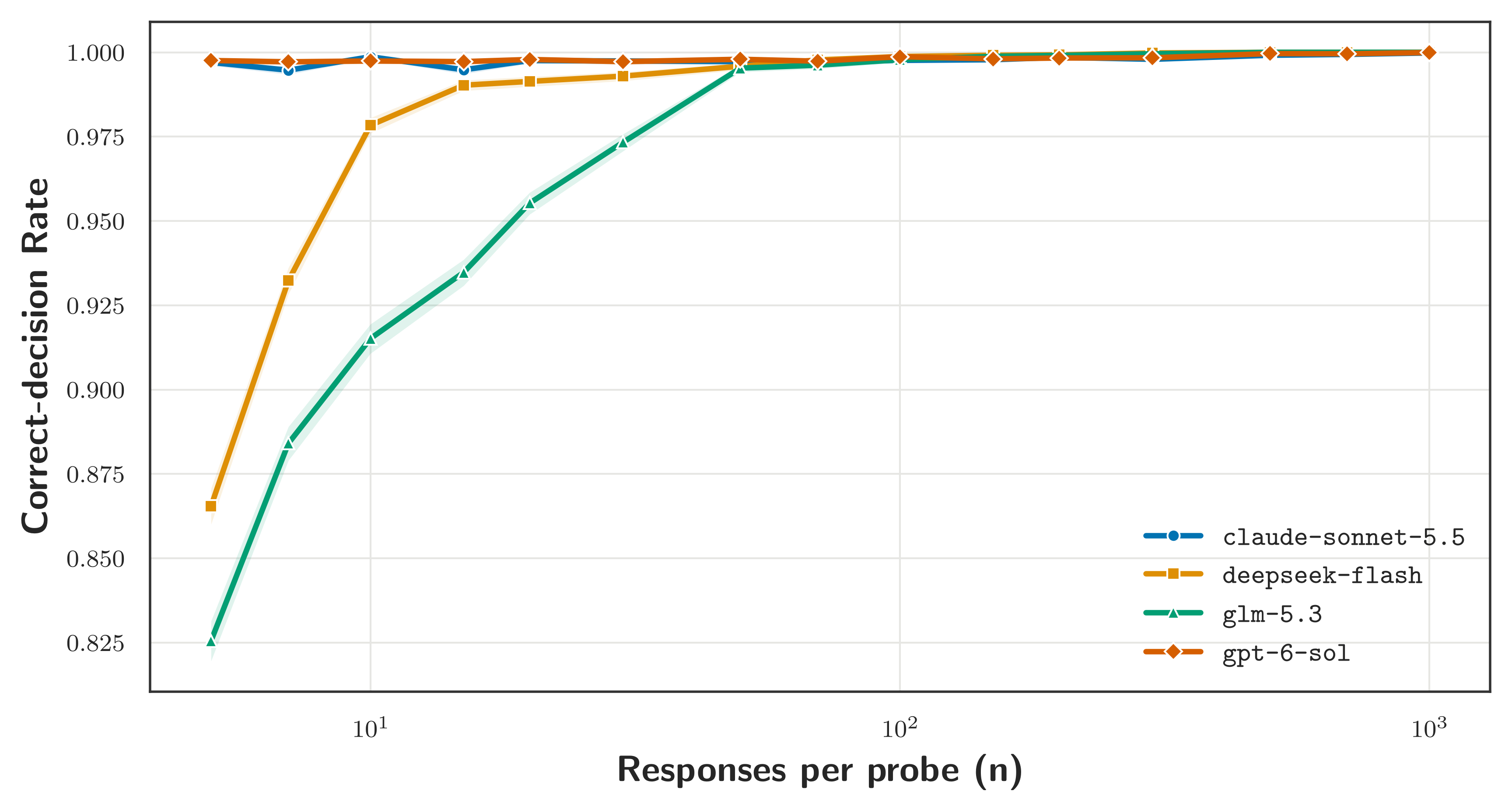}
  \end{minipage}
  \hfill
  \begin{minipage}[c]{0.43\linewidth}
    \centering
    \begin{tblr}{
  width = \linewidth,
  colspec = {X[2,halign=l,valign=m] X[1,halign=r,valign=m] X[1,halign=r,valign=m]},
  colsep = 3pt,
  rowhead = 1,
  rows = {font=\small},
  row{1} = {font=\small\bfseries},
}
\toprule
Model & Audit Cost & Fingerprint Cost \\
\midrule
% \texttt{gpt-6-luna} & $\mathdollar$0.0024 & $\mathdollar$0.041 \\
\texttt{claude-sonnet-5.5} & $\mathdollar$0.048 & $\mathdollar$0.960 \\
\texttt{deepseek-flash} & $\mathdollar$0.003 & $\mathdollar$0.060 \\
% \texttt{claude-opus-5.5} & $\mathdollar$0.084 & $\mathdollar$1.680 \\
\texttt{glm-5.3} & $\mathdollar$0.014 & $\mathdollar$0.280 \\
\texttt{gpt-6-sol} & $\mathdollar$0.035 & $\mathdollar$0.690 \\
\bottomrule
\end{tblr}

  \end{minipage}
  \caption{
    Detection success and monetary cost for four claimed models.
    \textbf{Left}: correct-decision rates at $\alpha = 0.05$ as the number of per-probe responses increases, averaged equally over the 16 serving models using 1,000 audits per claimed–serving pair and budget.
    \textbf{Right}: estimated costs in USD for one audit with 250 requests and one reference fingerprint with 5,000 requests
  }
  \label{fig:exp_cost_budget}
\end{figure}

The left panel of \Cref{fig:exp_cost_budget} reports detection success for \texttt{claude-sonnet-5.5}, \texttt{gpt-6-sol}, \texttt{glm-5.3}, and \texttt{deepseek-flash}.
\texttt{Claude} and \texttt{GPT} achieve success rates close to one even at the smallest evaluated budgets.
\texttt{DeepSeek} and \texttt{GLM} benefit more from additional responses: at $n = 10$, their success rates are approximately $0.978$ and $0.915$, respectively.
At $n = 50$, all four models exceed $0.995$, with smaller gains as the budget increases further.
For these four claimed models, 50 responses per probe capture most of the observed improvement in detection success.

\paragraph{Monetary cost.}

The right panel of \Cref{fig:exp_cost_budget} reports the estimated costs of an audit with $250$ requests and an reference fingerprint construction with $5{,}000$ requests.
Across the four models, the per-audit cost ranges from USD $0.003$--$0.048$, and one reference fingerprint construction costs USD $0.060$--$0.960$.
Reusing the reference fingerprint spreads this initial collection cost across subsequent audits.

\subsection{\ref{rq:3}: Robustness to Configuration Overrides}
\label{sec:exp_robustness}

A relay may modify the requested deployment configuration, potentially altering the response distribution while continuing to serve the claimed model. 
We examine how configuration overrides affect endpoint-consistency decisions when the relay continues to serve the claimed model.

\paragraph{Override protocol.}

We use the four models reported in~\ref{rq:2} and consider 26 configuration overrides spanning three severity levels.
For each claimed model $M_0$, we retain its reference fingerprint collected under baseline configuration C0 and apply the overrides to a suspect endpoint that continues to serve $M_0$.
Minor overrides introduce neutral system prompts, modest decoding changes, or shorter output limits.
Moderate overrides constrain response style, request uniform selection, or require JSON output.
Large overrides introduce persona or context changes, positional preferences, or nearly greedy decoding, including a configuration that combines long context with low temperature.
\tabref{tab:exp3_override_protocol} summarizes the configurations and their model coverage, and Appendix~B.3 provides the exact prompt texts.

\begin{table}[t]
\centering
\caption{Configuration overrides for the robustness experiment. C0 specifies no system prompt, temperature $1.0$, $\texttt{top\_p}=0.8$, and a 1,024-token completion limit where these controls are available.}
\label{tab:exp3_override_protocol}
\begin{tblr}{
  width = \linewidth,
  colspec = {Q[wd=1.35cm,halign=l,valign=m] Q[wd=1.6cm,halign=l,valign=m] X[halign=l,valign=m] Q[wd=2.6cm,halign=c,valign=m] Q[wd=1.8cm,halign=c,valign=m] Q[wd=1.2cm,halign=c,valign=m] Q[wd=1.5cm,halign=c,valign=m]},
  colsep = 4pt,
  rowsep = 3pt,
  rows = {font=\small},
  row{1} = {font=\small\bfseries},
}
\toprule
Severity & ID & Configuration & Temp. & \texttt{top\_p} & Max. tokens & Models \\
\midrule
\SetCell[r=5]{l,m} Minor & C1 & Neutral system prompt & $1.0$ & $0.8$ & 1,024 & All \\
 & C2 & Helpful assistant & $1.0$ & $0.8$ & 1,024 & All \\
 & C3--C6 & Temperature sweep & $0.8,\,0.9,\,1.1,\,1.2$ & $0.8$ & 1,024 & Open \\
 & C7--C9 & Top-$p$ sweep & $1.0$ & $0.7,\,0.9,\,1.0$ & 1,024 & Open \\
 & C10--C11 & Output limit & $1.0$ & $0.8$ & 16, 32 & All \\
\midrule
\SetCell[r=4]{l,m} Moderate & C12 & Direct answers & $1.0$ & $0.8$ & 1,024 & All \\
 & C13 & Lowercase answers & $1.0$ & $0.8$ & 1,024 & All \\
 & C14 & Uniform-choice guidance & $1.0$ & $0.8$ & 1,024 & All \\
 & C15 & JSON answer format & $1.0$ & $0.8$ & 1,024 & All \\
\midrule
\SetCell[r=8]{l,m} Large & C16 & Neutral persona & $1.0$ & $0.8$ & 1,024 & All \\
 & C17 & Long neutral context & $1.0$ & $0.8$ & 1,024 & All \\
 & C18 & Always first option & $1.0$ & $0.8$ & 1,024 & All \\
 & C19 & Prefer last option & $1.0$ & $0.8$ & 1,024 & All \\
 & C20 & Prefer middle option & $1.0$ & $0.8$ & 1,024 & All \\
 & C21--C23 & Low temperature & $0.0,\,0.1,\,0.2$ & $0.8$ & 1,024 & Open \\
 & C24--C25 & Restricted top-$p$ & $1.0$ & $0.1,\,0.2$ & 1,024 & Open \\
 & C26 & Long context + low temperature & $0.1$ & $0.8$ & 1,024 & Open \\
\bottomrule
\end{tblr}
\par\smallskip
{\footnotesize\raggedright Comma-separated values denote separate configurations, ordered by ID. C1--C26 comprise 11 minor, four moderate, and 11 large overrides; C26 is the only combined override. Models: \emph{Open} denotes \texttt{glm-5.3} and \texttt{deepseek-flash}. Model coverage reflects the available collected configurations. All configurations use the fixed five-prompt panel; decoding values apply where supported. Exact prompt texts are provided in Appendix~B.3.\par}
\end{table}

Each audit collects 50 responses per probe under the specified override and compares them with the fixed reference fingerprint.
We conduct 100 audits per model–override pair using Algorithm~\ref{alg:distributional_audit} and report the non-rejection rate.

\begin{figure}[t]
  \centering
  \includegraphics[width=0.8\linewidth]{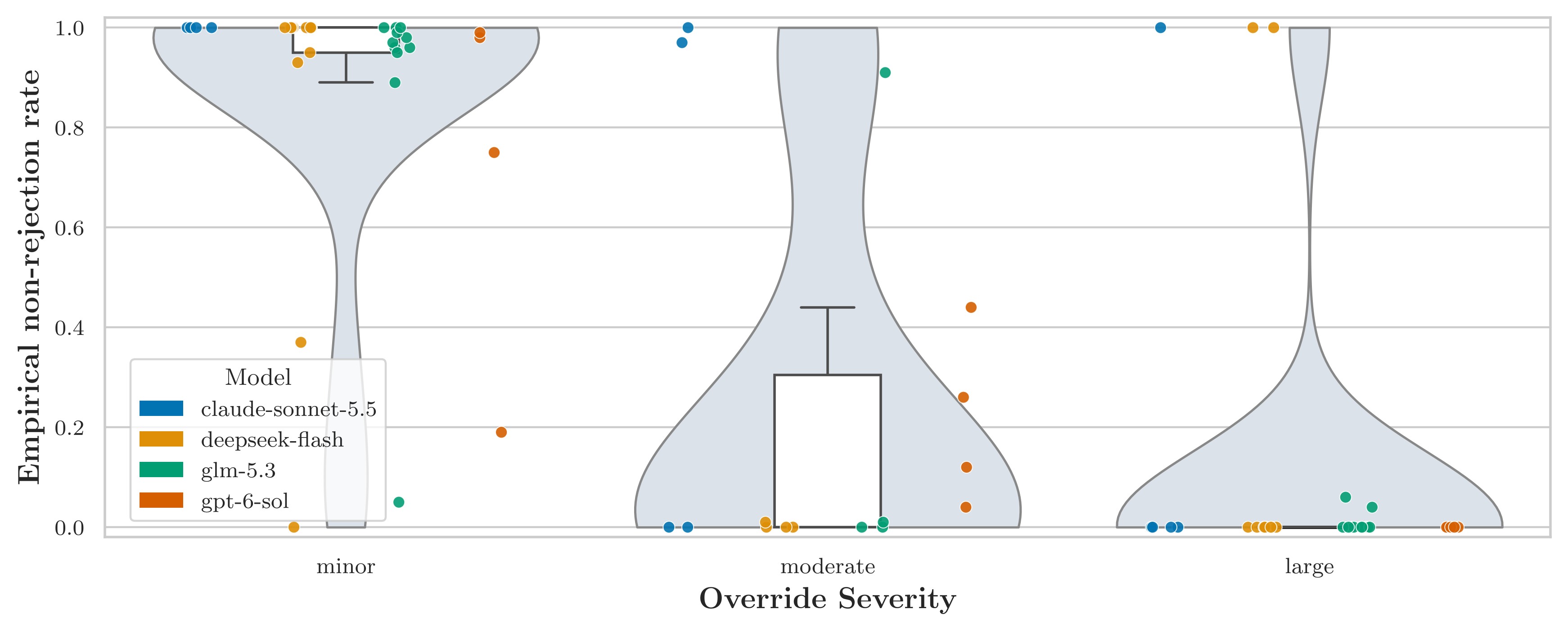}
  \caption{
    Empirical non-rejection rates under configuration overrides, grouped by severity and colored by model.
    Each point summarizes 100 audits for one model–override pair, using 50 responses per probe.
    Each audit compares the overridden endpoint with the stored reference fingerprint of the claimed model at $\alpha = 0.05$.
  }
  \label{fig:exp3_claim_consistency}
\end{figure}

\Cref{fig:exp3_claim_consistency} shows that empirical non-rejection rates generally decrease with override severity, with substantial variation across models and configurations.
Under minor overrides, \texttt{Claude}’s rates remain at or above 69\%, whereas some configurations produce frequent rejections for \texttt{GPT}, \texttt{GLM}, and \texttt{DeepSeek}. Moderate overrides yield uniformly low non-rejection rates for \texttt{DeepSeek} (0--1\%), while \texttt{Claude} and \texttt{GLM} exhibit wider variation across configurations.
These results indicate that even overrides categorized as minor can produce detectable departures from the reference fingerprint.

Under large overrides, non-rejection rates are 0\% for \texttt{GPT-6-Sol} across all evaluated configurations and range from 0--6\% for \texttt{GLM-5.3}.
\texttt{Claude Sonnet 5.5} and \texttt{DeepSeek Flash} also exhibit 0\% non-rejection rates under most large overrides, but both retain rates of 100\% under some configurations.
The severity groups therefore capture a broad trend, while the effect of an individual override depends on both the model and the specific configuration change.

\subsection{\ref{rq:4} \& \ref{rq:5}: Model Attribution and Substitution-Ratio Estimation}
\label{sec:model_attribution_and_mix_routing}

We simulate two relay scenarios to evaluate the likelihood-based inference framework in \secref{sec:method_inference}: full substitution and constant-ratio mixed routing.
Both experiments use the same $16$ models in \tabref{tab:model_inventory}.
Each model serves in turn as the claimed model, with the remaining $15$ forming the candidate substitute set, this gives $240$ ordered claimed–substitute pairs.

\begin{figure}[htbp]
  \centering
  \includegraphics[width=0.8\linewidth]{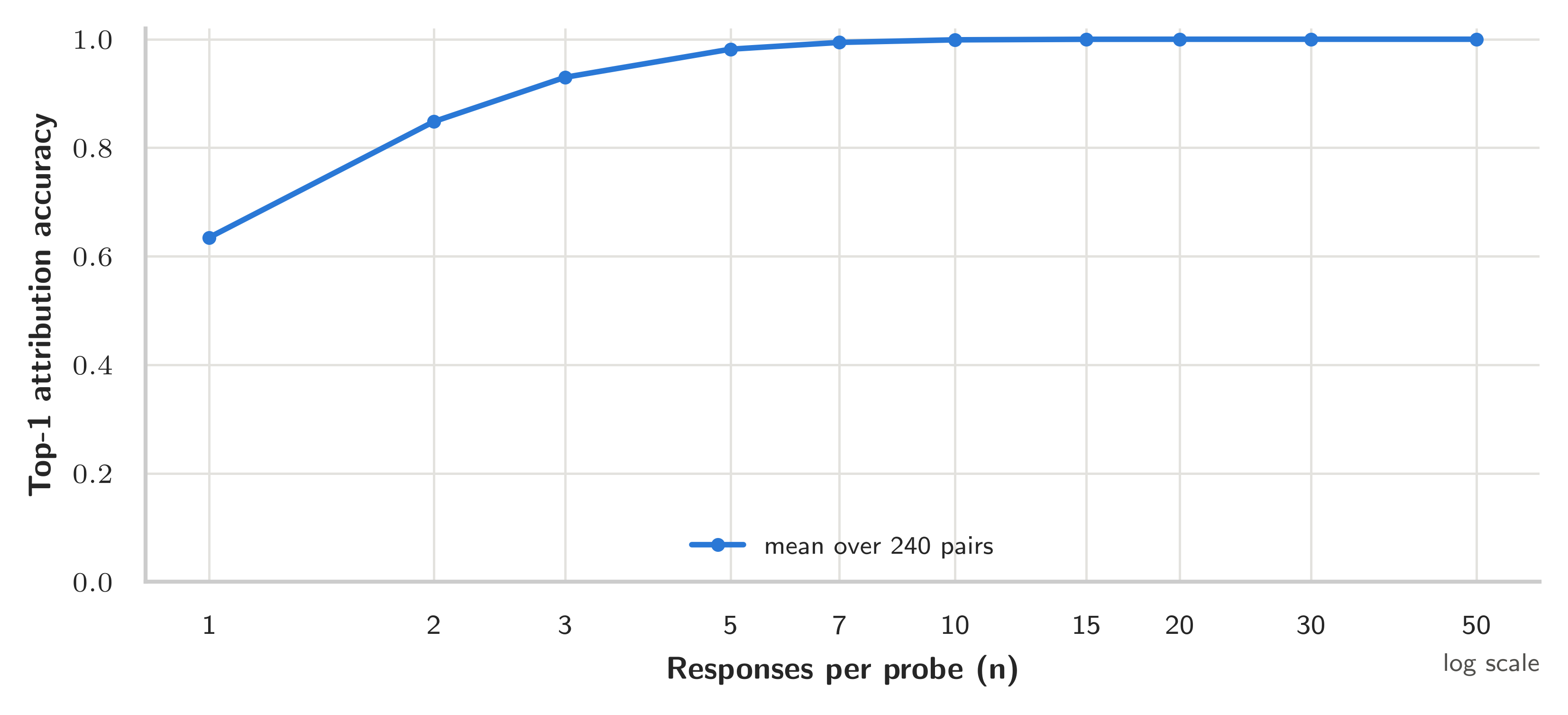}
  \caption{Top-1 attribution accuracy under full substitution ($\pi = 1$) versus responses per probe.
    The curve shows accuracy averaged equally over the 240 ordered claimed–substitute pairs.
    The horizontal axis uses a logarithmic scale.}
  \label{fig:exp4_accuracy_vs_n}
\end{figure}

\paragraph{Full substitution protocol.}

For each pair $(M_0, M_{\mathrm{sub}})$ with $M_{\mathrm{sub}} \neq M_0$, the simulated relay claims to serve $M_0$ but serves every request using $M_{\mathrm{sub}}$.
For each of the five probes, we draw a category-count vector from a multinomial distribution with $n$ trials and category probabilities given by the empirical fingerprint of $M_{\mathrm{sub}}$.
This is distributionally equivalent to resampling its stored responses to that probe with replacement and counting their response categories.
We apply Algorithm~\ref{alg:likelihood_inference} with the substitution ratio fixed at $\pi = 1$, vary the audit budget, and run 1,000 independent simulated audits per pair and budget.

\Cref{fig:exp4_accuracy_vs_n} shows that mean top-1 attribution accuracy increases rapidly with the audit budget.
Accuracy rises from $0.63$ with one response per probe to $0.98$ with five and $0.999$ with ten, reaching 1.00 at 30 and 50 responses per probe.
Within the evaluated candidate set, full-substitution attribution therefore achieves high accuracy at budgets below the 50 responses per probe used in~\ref{rq:1}.

\paragraph{Mixed-routing protocol.}

We use the same 240 ordered pairs.
The simulated relay claims to serve $M_0$, while each request is independently served by $M_{\mathrm{sub}}$ with probability $\pi$ and by $M_0$ with probability $1 - \pi$.
The substitution ratio $\pi$ is constant across probes, with $\pi = 0$ representing honest serving and $\pi = 1$ full substitution.
For each probe, we generate a multinomial category-count vector with $n$ trials using the mixture of the two empirical fingerprints, weighted by $1 - \pi$ for $M_0$ and $\pi$ for $M_{\mathrm{sub}}$.
This produces the same count distribution as independently routing each request, resampling a stored response from the selected model for that probe, and tallying the resulting categories.
We vary the audit budget and substitution ratio, running 200 independent simulated audits per pair, ratio, and budget.
The auditor is not given the substitute identity and jointly estimates it with the substitution ratio using \Cref{eq:joint_mixture_estimation}.

\begin{figure}[htbp]
  \centering
  \includegraphics[width=0.8\linewidth]{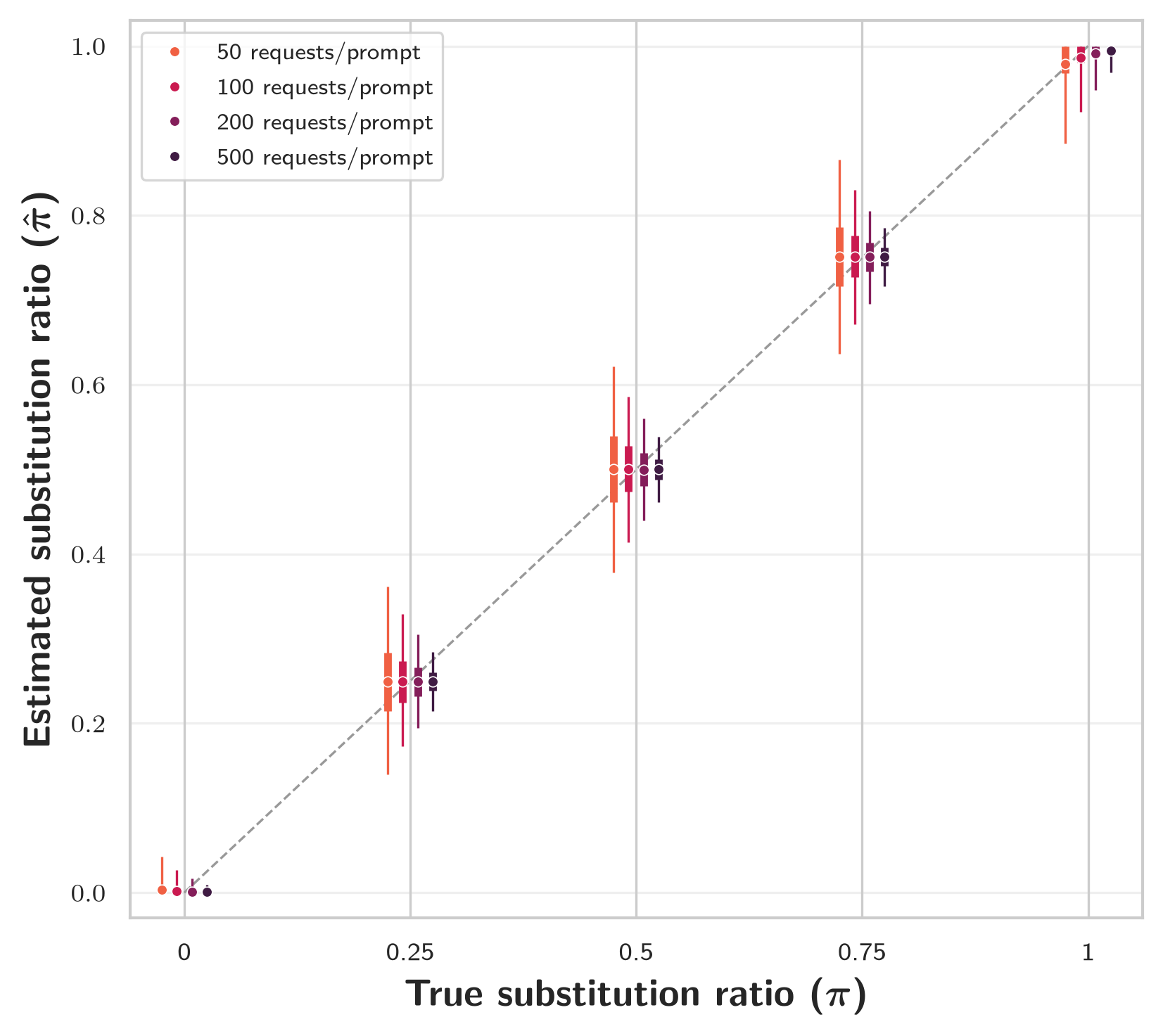}
  \caption{
    Estimated substitution ratio $\widehat{\pi}$ versus true substitution ratio $\pi$ under constant-ratio mixed routing at different audit budgets.
    Dots show means, thick bars show interquartile ranges, and thin bars show the 2.5th–97.5th percentile ranges across simulated audits pooled over 240 ordered claimed–substitute pairs.
    The dashed line indicates $\widehat{\pi} = \pi$.
  }
  \label{fig:exp5_pi_hat}
\end{figure}

\Cref{fig:exp5_pi_hat} shows that the mean estimated substitution ratio remains close to the true ratio across the evaluated settings.
The interquartile and 2.5--97.5\% ranges narrow as the budget increases from 50 to 500 responses per probe, indicating more precise estimates with additional observations.
Variation is largest near an equal mixture and smaller near the endpoints, where small inward deviations are consistent with constraining the estimate to $[0,1]$.
The estimates thus track the substitution ratio without being given the substitute identity.

\begin{figure}[htbp]
  \centering
  \includegraphics[width=0.8\linewidth]{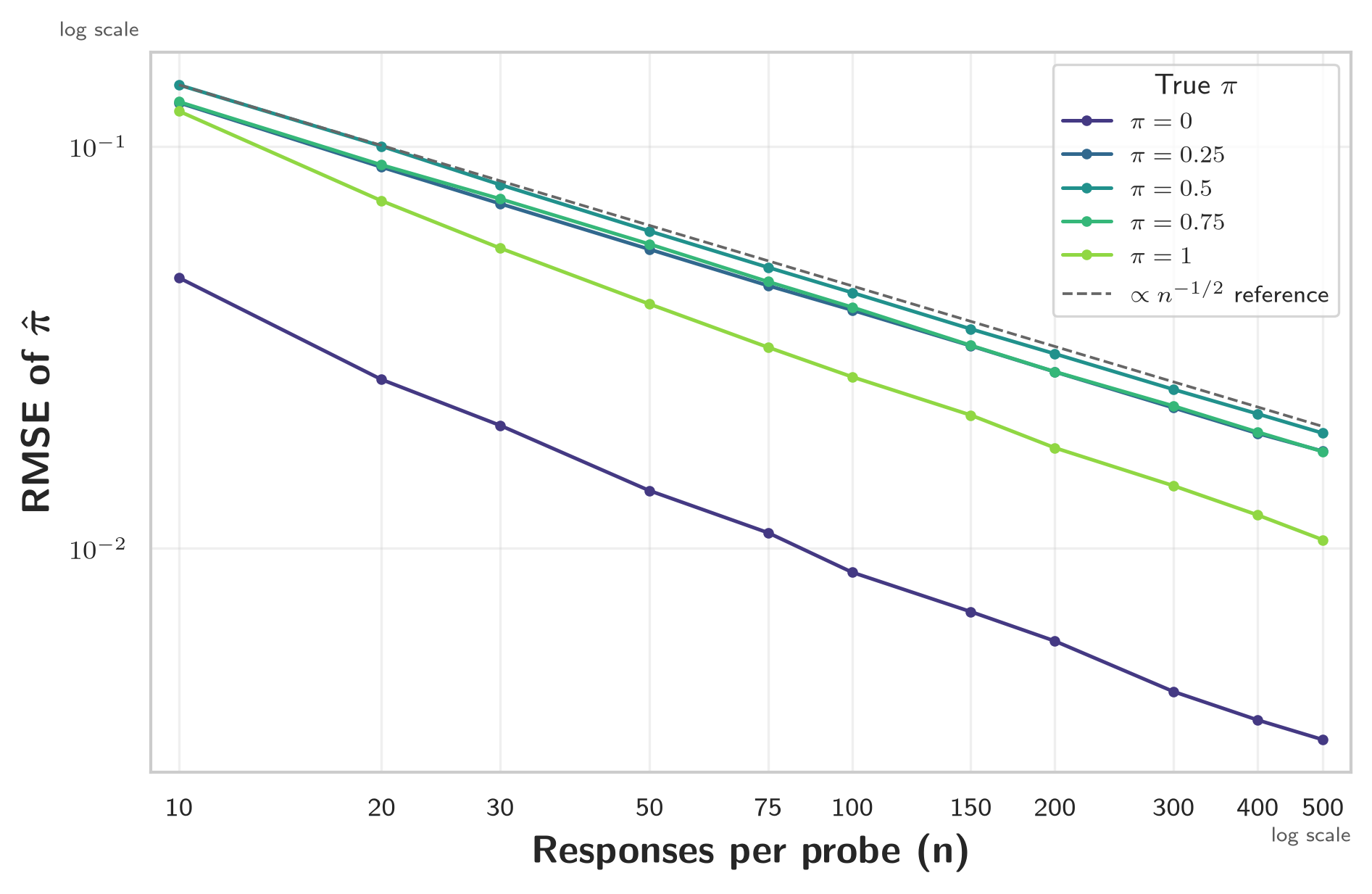}
  \caption{
    RMSE of the estimated substitution ratio versus responses per probe, $n$, under constant-ratio mixed routing.
    Each curve corresponds to a true substitution ratio $\pi$, with errors pooled over 240 ordered claimed–substitute pairs.
    Both axes use logarithmic scales, and the dashed line indicates the $n^{-1/2}$ reference rate.
}
  \label{fig:exp5_rmse}
\end{figure}

\paragraph{Estimation error versus budget.}

\Cref{fig:exp5_rmse} shows a steady decline in RMSE as the audit budget grows.
Among the displayed settings, error is highest at $\pi = 0.5$, decreasing from approximately 0.061 at 50 responses per probe to 0.030 at 200, whereas the honest-serving and full-substitution controls have smaller errors.
The interior-mixture curves approximately follow the $n^{-1/2}$ reference rate, consistent with the reduction in audit sampling variability.
With fixed fingerprints and a reliably identified substitute, increasing the audit budget reduces the variance of the estimated substitution ratio approximately in proportion to $1/n$, yielding the observed RMSE scaling when sampling variability dominates estimation error.
All displayed curves fall below $0.02$ at 500 responses per probe.
Substitution-ratio estimation therefore continues to benefit from additional observations after full-substitution attribution has reached near-perfect accuracy.

\section{Conclusion}\label{sec:conclu}

We introduced distributional fingerprint auditing (DFA) for assessing whether an LLM relay behaves consistently with its advertised model using only black-box query access. DFA constructs categorical fingerprints from repeated responses to simple finite-choice probes and combines probe-specific evidence into an approximate endpoint-level \(p\)-value. The registered fingerprints also support likelihood-based model attribution and substitution-ratio estimation. Controlled experiments on 16 proprietary and open-weight LLMs show clear separation between honest serving and full substitution, including within-family substitutions, using five probes and 250 relay requests per audit. Reusing reference fingerprints keeps subsequent audits inexpensive. Simulations based on stored empirical fingerprints further demonstrate accurate attribution under full substitution and increasingly precise substitution-ratio estimates under constant-ratio mixed routing as the audit budget grows.
The configuration experiments show that overrides can produce detectable departures from the reference fingerprint even when the served model remains unchanged, with effects varying across models and configurations. The practical scope of these audits depends on trusted reference access and on whether audited traffic represents ordinary service behavior. Evaluating inference on live mixed-routing endpoints, accounting for uncertainty in reference fingerprints, and addressing relays that selectively recognize audit requests are important directions for extending DFA to less controlled deployments.

% \section*{Acknowledgments}
% This was supported in part by......

\newpage
%Bibliography
\bibliographystyle{unsrt}  
\bibliography{references}  

@article{sun2025invisible,
  title={Invisible tokens, visible bills: The urgent need to audit hidden operations in opaque llm services},
  author={Sun, Guoheng and Wang, Ziyao and Zhao, Xuandong and Tian, Bowei and Shen, Zheyu and He, Yexiao and Xing, Jinming and Li, Ang},
  journal={arXiv preprint arXiv:2505.18471},
  year={2025}
}

@article{zhang2026real,
  title={Real money, fake models: Deceptive model claims in shadow APIs},
  author={Zhang, Yage and Jiang, Yukun and Chen, Zeyuan and Backes, Michael and Shen, Xinyue and Zhang, Yang},
  journal={arXiv preprint arXiv:2603.01919},
  year={2026}
}

@article{egashira2024exploiting,
  title={Exploiting llm quantization},
  author={Egashira, Kazuki and Vero, Mark and Staab, Robin and He, Jingxuan and Vechev, Martin},
  journal={Advances in Neural Information Processing Systems},
  volume={37},
  pages={41709--41732},
  year={2024}
}

@inproceedings{shen2024anything,
  title={" do anything now": Characterizing and evaluating in-the-wild jailbreak prompts on large language models},
  author={Shen, Xinyue and Chen, Zeyuan and Backes, Michael and Shen, Yun and Zhang, Yang},
  booktitle={Proceedings of the 2024 on ACM SIGSAC Conference on Computer and Communications Security},
  pages={1671--1685},
  year={2024}
}

@misc{vercel2026gateway,
  title        = {Vercel AI Gateway Documentation},
  howpublished = {\url{https://vercel.com/docs/ai-gateway}}
}

@misc{openrouter2026quickstart,
  title        = {OpenRouter Quickstart},
  howpublished = {\url{https://openrouter.ai/docs}}
}

@misc{litellm2026routing,
  title        = {LiteLLM Routing and Fallback Documentation},
  howpublished = {\url{https://docs.litellm.ai/docs}}
}

@article{velasco2025your,
  title={Is your LLM overcharging you? tokenization, transparency, and incentives},
  author={Velasco, Ander Artola and Tsirtsis, Stratis and Okati, Nastaran and Gomez-Rodriguez, Manuel},
  journal={arXiv preprint arXiv:2505.21627},
  year={2025}
}

@article{shao2025sok,
  title={Sok: Large language model copyright auditing via fingerprinting},
  author={Shao, Shuo and Li, Yiming and He, Yu and Yao, Hongwei and Yang, Wenyuan and Tao, Dacheng and Qin, Zhan},
  journal={arXiv preprint arXiv:2508.19843},
  year={2025}
}

@inproceedings{pasquini2025llmmap,
  title={$\{$LLMmap$\}$: Fingerprinting for large language models},
  author={Pasquini, Dario and Kornaropoulos, Evgenios M and Ateniese, Giuseppe},
  booktitle={34th USENIX Security Symposium (USENIX Security 25)},
  pages={299--318},
  year={2025}
}

@article{sun2025idiosyncrasies,
  title={Idiosyncrasies in large language models},
  author={Sun, Mingjie and Yin, Yida and Xu, Zhiqiu and Kolter, J Zico and Liu, Zhuang},
  journal={arXiv preprint arXiv:2502.12150},
  year={2025}
}

@inproceedings{met,
  title={Model equality testing: Which model is this api serving?},
  author={Gao, Irena and Liang, Percy and Guestrin, Carlos},
  booktitle={International Conference on Learning Representations},
  volume={2025},
  pages={86369--86382},
  year={2025}
}

@article{kbf,
  title={KBF: Knowledge Boundary as Fingerprint for Language Model and Black-Box API Auditing},
  author={Fang, Yijia and Feng, Yiqing and Li, Bingyu and Zhou, Mingxun},
  journal={arXiv preprint arXiv:2605.29524},
  year={2026}
}

@article{tee,
  title={Are you getting what you pay for? auditing model substitution in llm apis},
  author={Cai, Will and Shi, Tianneng and Zhao, Xuandong and Song, Dawn},
  journal={arXiv preprint arXiv:2504.04715},
  year={2025}
}

@inproceedings{ren2024robust,
  title={A robust semantics-based watermark for large language model against paraphrasing},
  author={Ren, Jie and Xu, Han and Liu, Yiding and Cui, Yingqian and Wang, Shuaiqiang and Yin, Dawei and Tang, Jiliang},
  booktitle={Findings of the Association for Computational Linguistics: NAACL 2024},
  pages={613--625},
  year={2024}
}

@article{shao2026fit,
  title={Fit-print: Towards false-claim-resistant model ownership verification via targeted fingerprint},
  author={Shao, Shuo and Zhu, Haozhe and Li, Yiming and Yao, Hongwei and Zhang, Tianwei and Qin, Zhan},
  journal={IEEE Transactions on Information Forensics and Security},
  year={2026},
  publisher={IEEE}
}

@inproceedings{xu2024instructional,
  title={Instructional Fingerprinting of Large Language Models},
  author={Xu, Jiashu and Wang, Fei and Ma, Mingyu and Koh, Pang Wei and Xiao, Chaowei and Chen, Muhao},
  booktitle={Proceedings of the 2024 Conference of the North American Chapter of the Association for Computational Linguistics: Human Language Technologies (Volume 1: Long Papers)},
  pages={3277--3306},
  year={2024},
  publisher={Association for Computational Linguistics},
  doi={10.18653/v1/2024.naacl-long.180}
}

@inproceedings{kirchenbauer2023watermark,
  title={A Watermark for Large Language Models},
  author={Kirchenbauer, John and Geiping, Jonas and Wen, Yuxin and Katz, Jonathan and Miers, Ian and Goldstein, Tom},
  booktitle={Proceedings of the 40th International Conference on Machine Learning},
  series={Proceedings of Machine Learning Research},
  volume={202},
  pages={17061--17084},
  year={2023},
  publisher={PMLR}
}

@article{shao2025zeroprint,
  title={Reading Between the Lines: Towards Reliable Black-box LLM Fingerprinting via Zeroth-order Gradient Estimation},
  author={Shao, Shuo and Li, Yiming and Yao, Hongwei and Chen, Yifei and Yang, Yuchen and Qin, Zhan},
  journal={arXiv preprint arXiv:2510.06605},
  year={2025}
}

@inproceedings{gubri2024trap,
  title={Trap: Targeted random adversarial prompt honeypot for black-box identification},
  author={Gubri, Martin and Ulmer, Dennis and Lee, Hwaran and Yun, Sangdoo and Oh, Seong Joon},
  booktitle={Findings of the Association for Computational Linguistics: ACL 2024},
  pages={11496--11517},
  year={2024}
}

@article{bruckner2026onetoken,
  title={One Token Is Enough: Fingerprinting and Verifying Large Language Models from Single-Token Output Distributions},
  author={Br{\"u}ckner, Tom{\'a}{\v{s}}},
  journal={arXiv preprint arXiv:2607.10252},
  year={2026}
}

@inproceedings{zhu2026rut,
  title={Auditing Black-Box LLM APIs with a Rank-Based Uniformity Test},
  author={Zhu, Xiaoyuan and Ye, Yaowen and Qiu, Tianyi and Zhu, Hanlin and Tan, Sijun and Mannan, Ajraf and Michala, Jonathan and Popa, Raluca Ada and Neiswanger, Willie},
  booktitle={The Fourteenth International Conference on Learning Representations},
  year={2026},
  url={https://openreview.net/forum?id=PSIe9mmF7a}
}

@inproceedings{chen2022modelchange,
  title={How Did the Model Change? Efficiently Assessing Machine Learning API Shifts},
  author={Chen, Lingjiao and Zaharia, Matei and Zou, James},
  booktitle={The Tenth International Conference on Learning Representations},
  year={2022},
  url={https://openreview.net/forum?id=gFDFKC4gHL4}
}

@article{chen2023chatgpt,
  title={How Is {ChatGPT}'s Behavior Changing over Time?},
  author={Chen, Lingjiao and Zaharia, Matei and Zou, James},
  journal={arXiv preprint arXiv:2307.09009},
  year={2023}
}

@inproceedings{eyuboglu2024changelists,
  title={Model {ChangeLists}: Characterizing Updates to {ML} Models},
  author={Eyuboglu, Sabri and Goel, Karan and Desai, Arjun and Chen, Lingjiao and Monfort, Mathew and R{\'e}, Christopher and Zou, James},
  booktitle={Proceedings of the 2024 ACM Conference on Fairness, Accountability, and Transparency},
  pages={2432--2453},
  year={2024},
  publisher={Association for Computing Machinery},
  doi={10.1145/3630106.3659047}
}

@article{pearson1900criterion,
  title={X. On the Criterion that a Given System of Deviations from the Probable in the Case of a Correlated System of Variables Is Such that It Can Be Reasonably Supposed to Have Arisen from Random Sampling},
  author={Pearson, Karl},
  journal={The London, Edinburgh, and Dublin Philosophical Magazine and Journal of Science},
  volume={50},
  number={302},
  pages={157--175},
  year={1900},
  doi={10.1080/14786440009463897}
}

@article{liu2019acat,
  title={ACAT: a fast and powerful p value combination method for rare-variant analysis in sequencing studies},
  author={Liu, Yaowu and Chen, Sixing and Li, Zilin and Morrison, Alanna C and Boerwinkle, Eric and Lin, Xihong},
  journal={The american journal of human genetics},
  volume={104},
  number={3},
  pages={410--421},
  year={2019},
  publisher={Elsevier}
}

@article{cct,
  title={Cauchy combination test: a powerful test with analytic p-value calculation under arbitrary dependency structures},
  author={Liu, Yaowu and Xie, Jun},
  journal={Journal of the American Statistical Association},
  volume={115},
  number={529},
  pages={393--402},
  year={2020},
  publisher={Taylor \& Francis}
}

@inproceedings{renda2023sampling,
  title={Can {LLMs} Generate Random Numbers? Evaluating {LLM} Sampling in Controlled Domains},
  author={Renda, Alex and Hopkins, Aspen K. and Carbin, Michael},
  booktitle={ICML 2023 Workshop: Sampling and Optimization in Discrete Space},
  year={2023},
  url={https://people.csail.mit.edu/renda/llm-sampling-paper}
}

@article{coronado2025deterministic,
  title={Deterministic or Probabilistic? The Psychology of {LLMs} as Random Number Generators},
  author={Coronado-Bl{\'a}zquez, Javier},
  journal={arXiv preprint arXiv:2502.19965},
  year={2025},
  doi={10.48550/arXiv.2502.19965},
  url={https://arxiv.org/abs/2502.19965}
}

% \newpage
% \input{sec/x_appendix}

\end{document}